\documentclass[aps,prl,twocolumn,showpacs]{revtex4}
\usepackage{amsfonts,amsmath,amssymb,graphicx}
\usepackage{amsfonts}
\usepackage{amsmath}
\usepackage{amssymb}
\usepackage{graphicx}

\def\kbar{\protect\@kbar}
\def\@kbar{\relax \bgroup
\def\@tempa{\hbox{\raise.73\ht0
\hbox to0pt{\kern.25\wd0\vrule width.5\wd0 height.1pt
depth.1pt\hss}\box0}}\mathchoice{\setbox0\hbox{$\displaystyle
k$}\@tempa}{\setbox0\hbox{$\textstyle
k$}\@tempa}{\setbox0\hbox{$\scriptstyle
k$}\@tempa}{\setbox0\hbox{$\scriptscriptstyle k$}\@tempa}\egroup}

\begin{document}

\title{\textbf{Generic Regularities in Quasienergy Spectra and Free Quantum Dynamics Independently of Chaos Strength}}
\author{Itzhack Dana}
\affiliation{Minerva Center and Department of Physics, Bar-Ilan University, Ramat-Gan
52900, Israel}

\begin{abstract}
Generic and significant regularities are shown to occur in the quasienergy spectra of the generalized quantum kicked particle for arbitrary quasimomentum, a quantity most relevant in atom-optics experimental realizations of this paradigmatic nonintegrable system. The spectral regularities are basically different from the usual ones associated with ordered regions in a mixed phase space, since they are \emph{completely independent} of the chaos strength. Their origin are free-motion features unaffected by the nonintegrability. The generic regularities are \emph{dense} subspectra, each being either a regular level sequence with approximately a Poisson level-spacing distribution or a ladder of equally-spaced levels. The quantum dynamics associated with each regular subspectrum is essentially \emph{free} for \emph{arbitrarily strong} chaos.
\newline
\end{abstract}
\maketitle

The nature of the energy and quasienergy spectra of quantum systems whose
classical limit is nonintegrable has been the subject of an enormous number
of studies during the last four decades. Percival \cite{ip} made the
pioneering distinction between \textquotedblleft regular\textquotedblright\
and \textquotedblleft irregular\textquotedblright\ energy spectra of systems
which are classically integrable and completely chaotic, respectively. This
was followed by the first works \cite{bt,wd,br} studying the statistics of
energy levels, mainly the level-spacing distribution. For integrable
systems, this distribution is Poisson \cite{bt}, with level clustering,
while for completely chaotic systems it is Wigner \cite{wd}, with level
repulsion. For systems with a mixed phase space, the distribution
was shown to be, under some assumptions, a weighted superposition of Poisson
and Wigner distributions associated with the ordered and chaotic phase-space
regions \cite{br}.

For time-periodic quantum systems, the energy is replaced by the quasienergy
(QE), giving the eigenvalues of the one-period evolution operator.
Paradigmatic and realistic models are the kicked-rotor systems \cite{dl,al,ali,cs,ff,fmi,fm,fmir,kp,qr,e1,e2,e3,e4,e5,e6} exhibiting a variety of phenomena, the most well-known one being dynamical localization \cite{dl}, i.e., a quantum suppression of the classical chaotic diffusion for generic irrational values of a scaled Planck constant $\hbar _{\mathrm{s}}$. This phenomenon can be attributed to an Anderson-like localization of QE eigenstates in angular-momentum space 
\cite{al}. Numerically, the QE level-spacing distribution for the usual kicked rotor was found to be Poisson for moderate chaos (nonintegrability) strength \cite{ff} or in a quantum regime ($\hbar _{\mathrm{s}}\sim 1$) \cite{fm}. The distribution turns into Wigner for very strong chaos \cite{fmi} or in a semiclassical regime ($\hbar _{\mathrm{s}}\ll 1$) \cite{fm}.

During the last two decades, kicked-rotor systems have been
experimentally realized using atom-optics techniques with cold atoms or
Bose-Einstein condensates \cite{e1,e2,e3,e4,e5,e6}. This allowed to
observe in the laboratory several quantum-chaos phenomena, including
dynamical localization \cite{e1}, and to verify theoretical predictions. In
the experiments, the kicked rotor and variants of it were actually realized
as kicked-\emph{particle} systems, since atoms move on lines and not on
circles like rotors. These realizations are based on the fact that a kicked particle reduces to a generalized kicked rotor at any fixed value of the conserved particle \emph{quasimomentum} $\beta $ \cite{kp,qr} (see also below). The usual kicked rotor, whose QE spectral statistics has been studied (see above),
corresponds to the particular case of $\beta =0$. However, several important
phenomena arise in wide ranges of $\beta$ and have been experimentally
realized \cite{e2,e3,e4,e5}. In addition, a general wavepacket of the
quantum kicked particle is the superposition of QE states with \emph{all}
values of $\beta $ \cite{kp,qr}. It is thus natural to ask about the
regularity and irregularity properties of QE spectra for arbitrary $\beta $.

In this work, we show that the QE spectrum of the quantum kicked particle
for all $\beta $ \emph{generically} exhibits significant regularities that are basically different in nature from the usual ones \cite{br}, associated with classically ordered regions in a mixed phase space. In fact, the new kinds of spectral regularities are \emph{completely independent} of the chaos strength and therefore persist even in fully chaotic regimes. Most of our results are exact and hold for rather general kicked-particle systems. For definiteness, we consider here the generalized version of the ordinary quantum kicked particle, described in scaled variables by the Hamiltonian 
\begin{equation}
\hat{H}=\frac{\hat{p}^{2}}{2}+kV(\hat{x})\sum_{s=-\infty }^{\infty }\delta
(t-s),  \label{KP}
\end{equation}
where $\hat{x}$ and $\hat{p}$ are position and momentum operators ($[\hat{x},
\hat{p}]=i\hbar $), $k$ is a nonintegrability parameter, and $V(\hat{x})$ is
a general $2\pi $-periodic potential. We show that the total QE spectrum of (\ref{KP}) for all $\beta$ is the superposition of fully regular and generically \emph{dense} subspectra (\ref{wj}) that are essentially independent of the nonintegrability $kV(\hat{x})$. For irrational values of $\hbar _{\mathrm{s}}=\hbar /(2\pi )$, each subspectrum has approximately a Poisson level-spacing distribution, as illustrated in Fig. 1. For rational $\hbar _{\mathrm{s}}$, corresponding to quantum resonance \cite{fmir,kp,qr,e2,e3,note0}, a subspectrum at any fixed irrational value of $\beta $ is a ladder of equally-spaced levels covering densely all the QE range. If also $\beta $ is rational \cite{note0}, the number of levels is finite and practically \emph{no} ladder regularity occurs for $\beta =0$ (usual kicked rotor), see Fig. 2. A regular subspectrum originates from free-motion features unaffected by the nonintegrability. It is then shown that the quantum dynamics associated with it is essentially \emph{free} for \emph{arbitrarily strong} chaos.
 
\begin{figure}[tbp]
\includegraphics[width=8.7cm]{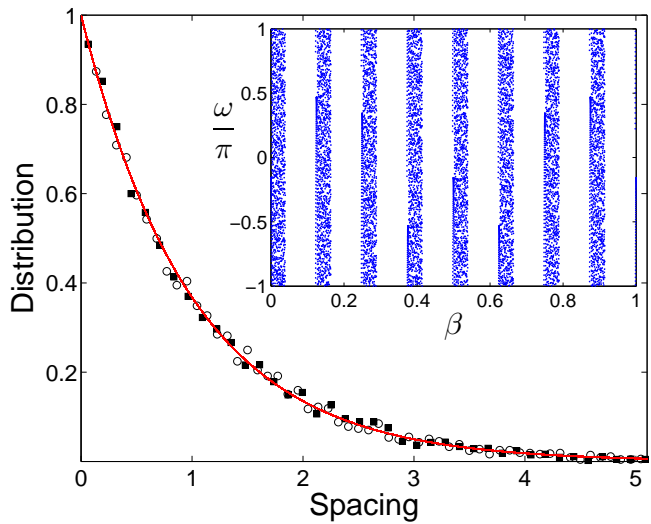}
\caption{(Color online) Distribution of the spacings $S$ between neighboring
QE levels of a regular subspectrum (\ref{wj}) associated with quasimomenta (\ref{bj}) ($j=1,...,10^{4}$) for $\beta =\sqrt{2}-1$
and: $\hbar _{\mathrm{s}}=\sigma =(\sqrt{5}+1)/2$ (open
circles), $\hbar _{\mathrm{s}}=(8+10^{-4}\sigma /3)/13$ (filled
squares). The solid (red) line is the Poisson distribution $P(S)=\exp (-S)$. The average spacing $\langle S\rangle $ is normalized to $1$.
The inset shows the levels $\omega_j$ and quasimomenta $\beta_j$ in the second case of $\hbar _{\mathrm{s}}$; the $10^{4}$ values of $\beta_j$ did not \textquotedblleft explore" yet all the $\beta$ range since $\hbar _{\mathrm{s}}$ is very close to a rational value ($8/13$). Still, the corresponding level-spacing
distribution agrees well with the Poisson one.}
\label{fig1}
\end{figure}
\begin{figure}[tbp]
\includegraphics[width=8.7cm]{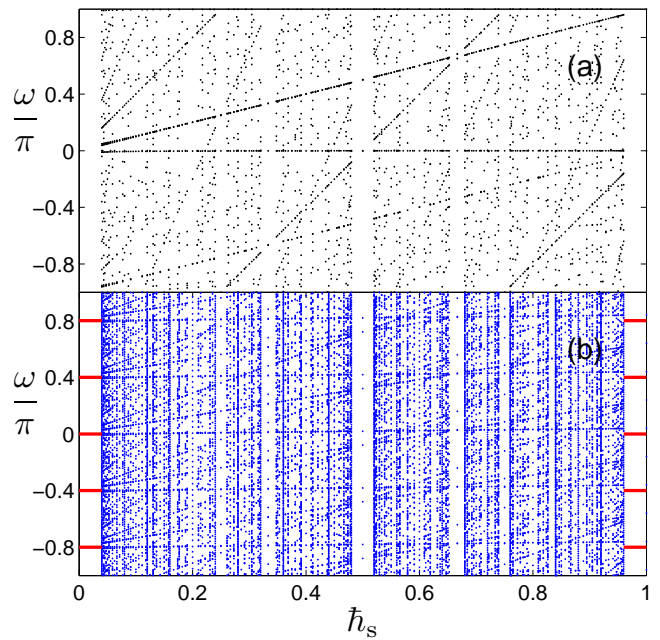}
\caption{(Color online) QE spectra of (\ref{KP}) as functions of $\hbar _{\mathrm{s}}=\hbar /(2\pi )$ for $k=0.1$, $V(x)=\cos (x)$, and: (a) $\beta =0$ (usual kicked rotor); (b) $\beta =0.2$, featuring spectral ladders with dominant spacing $\Delta \omega =2\pi /5$. These ladders are indicated by the left and right solid (red) segments. In both cases, $\hbar _{\mathrm{s}}$ takes all rational values in $[0,1)$ with denominators $\leq 25$.}
\label{fig2}
\end{figure}
First, some preliminaries. The one-period evolution operator for (\ref{KP}), from $t=s-0$ to $t=s+1-0$, is
\begin{equation}
\hat{U}=\exp [-i\hat{p}^{2}/(2\hbar )]\exp [-ikV(\hat{x})/\hbar ].  \label{U}
\end{equation}
The QE states $\Psi _{\omega }(x)$ with QE $\omega $ ($0\leq \omega <2\pi $)
are the eigenstates of (\ref{U}):
\begin{equation}
\hat{U}\Psi _{\omega }(x)=\exp (-i\omega )\Psi _{\omega }(x).  \label{qee}
\end{equation}
The $2\pi $-periodicity of (\ref{U}) in $\hat{x}$ implies that $\Psi
_{\omega }(x)$ can be chosen to have the Bloch form: 
\begin{equation}
\Psi _{\omega }(x)=\exp (i\beta x)\psi _{\beta ,\omega }(x),  \label{bf}
\end{equation}
where $\beta $ is the quasimomentum ($0\leq \beta <1$) or the
\textquotedblleft fractional" part of the momentum in units of $\hbar$ \cite{kp,note} and $\psi_{\beta ,\omega }(x)$ is $2\pi $-periodic in $x$. One can then interpret $\psi_{\beta ,\omega }(x)$ as the QE state of a \textquotedblleft $\beta $-kicked-rotor\textquotedblright\ with angle $\theta =x$. This provides the connection between the kicked particle and the ensemble of all $\beta$-kicked-rotors. See more details in Ref. \cite{kp} and in note \cite{note}.

Next, consider the operators $\hat{T}_{j}=\exp (2\pi ij\hat{x}/\hbar )$ for
all integers $j$. Denoting $\hbar _{\mathrm{s}}=\hbar /(2\pi )$, the application of $\hat{T}_{j}$ to the QE state
(\ref{bf}) gives the Bloch states
\begin{eqnarray}
\hat{T}_{j}\Psi _{\omega }(x) &=& \exp[i(\beta +j/\hbar _{\mathrm{s}})x]
\psi _{\beta ,\omega}(x) \notag \\ 
&=& \exp (i\beta_j x)\psi^{(j)} _{\beta ,\omega }(x),  \label{tqe}
\end{eqnarray}
having quasimomenta (fractional parts of $\beta +j/\hbar _{\mathrm{s}}$)
\begin{equation}
\beta _{j}=\beta + j/\hbar _{\mathrm{s}} \ \ \mathrm{mod}(1)  \label{bj}
\end{equation}
and $2\pi$-periodic parts 
\begin{equation}
\psi^{(j)}_{\beta ,\omega }(x)= \exp (in_j x)\psi_{\beta ,\omega }(x),
\label{ppj}
\end{equation}
where $n_j$ is the integer part of $\beta +j/\hbar _{\mathrm{s}}$.

Now, from $[\hat{x},\hat{p}]=i\hbar $ or $\hat{x}=i\hbar d/dp$, we see that 
$\hat{T}_{j}=\exp (2\pi ij\hat{x}/\hbar )$ is a translation $\exp (-2\pi
jd/dp)$ in momentum by $-2\pi j$. This implies that 
\begin{equation}
\hat{T}_{j}e^{-i\hat{p}^{2}/(2\hbar )}=e^{-i\hat{p}^{2}/(2\hbar )}e^{-2i\pi ^2
j^{2}/\hbar }e^{2\pi ij\hat{p}/\hbar }\hat{T}_{j}.
\label{iden}
\end{equation}
In turn, since $\hat{p}=-i\hbar d/dx$, $\exp (2\pi ij\hat{p}/\hbar )$ in Eq. (\ref{iden}) is a translation $\exp (2\pi jd/dx)$ in $x$ by $2\pi j$. Like $\hat{T}_{j}$, this translation obviously commutes with $V(\hat{x})$. Then, by applying $\hat{T}_{j}$ to both sides of (\ref{qee}), using (\ref{bf}) and (\ref{iden}), we easily find that a state (\ref{tqe}) is an
eigenstate of $\hat{U}$ with QE 
\begin{equation}
\omega _{j}=\omega +2\pi j\beta +\pi j^{2}/\hbar _{\mathrm{s}} \ \ 
\mathrm{mod}(2\pi ).  \label{wj}
\end{equation}
The sequence (\ref{wj}) is fully regular and, apart of the constant term $\omega$, is independent of the nonintegrability, since it reflects free-motion features [Eq. (\ref{iden})] unaffected by a periodic potential $V(\hat{x})$. This has an analogue in the classical map for (\ref{KP}), $p_{s+1}=p_s+kf(x_s)$, $x_{s+1}=x_s+p_{s+1}$ [$f(x)=-dV/dx$]. Given an orbit $(x_s,p_s)$ with total energy $E_s=p_s^2/2+kV(x_s)$, it is easy to show that, for any integer $j$ and some integers $m_s$, also $(x_s+2\pi m_s,p_s+2\pi j)$ is an orbit whose total energy is $E_{j,s}=E_s+\Delta E_{j,s}$, where $\Delta E_{j,s}=2\pi jp_s+2\pi^2j^2$ is independent of $V(x_s)$. 

The QE levels (\ref{wj}) are associated with the fixed set of quasimomenta (\ref{bj}) and form a subspectrum of the total QE spectrum for all $\beta$. The latter is the superposition of the subspectra (\ref{wj}) for all initial $(\beta, \omega )$. If $\hbar _{\mathrm{s}}$ is irrational, $\omega _j$ (resp. $\beta _{j}$) will fill densely the entire $\omega$ (resp. $\beta$) range as $j\rightarrow \infty $, see the inset of Fig. 1. The levels (\ref{wj}) then form a dense subspectrum of the total spectrum. In addition, numerical studies \cite{fmir} of sequences similar to (\ref{wj}) have established that, generically, the distribution of spacings between neighboring elements is approximately Poisson. We have also extensively confirmed this, see Fig. 1.

A state (\ref{ppj}) is an eigenstate of the $\beta_j$-kicked-rotor with QE (\ref{wj}) and is just a translation of the $j=0$ state by $-n_j\hbar$ in angular momentum $n\hbar$ ($n$ integer). For generic irrational $\hbar _{\mathrm{s}}$, the state (\ref{ppj}) should feature dynamical localization in angular momentum, as for $\beta=0$ \cite{al}. Its localization center is located at $(\bar{n}-n_j)\hbar$, where $\bar{n}\hbar$ is the localization center for $j=0$. The states (\ref{ppj}) may look very correlated when they strongly overlap in angular-momentum space due to a large localization length in fully chaotic regimes. This does not contradict the fact that the corresponding level-spacing distribution is always approximately Poisson since the levels (\ref{wj}) are associated with different quasimomenta (\ref{bj}). 

The quantum kicked particle exhibits the phenomenon of quantum resonance, i.e., a quadratic growth in time of the mean kinetic energy for rational values of both $\hbar _{\mathrm{s}}$ and $\beta$ \cite{kp,qr,e2,note0}. Then, if $\hbar _{\mathrm{s}}$ is rational but $\beta$ is irrational, one expects dynamical localization to occur and this indeed follows from analytical results in simple cases \cite{kp,qr} (see also below). Despite this, manifestations of quantum resonance are felt for rational $\hbar _{\mathrm{s}}$ and generic $\beta$ close to some rational value \cite{kp,qr} and have been experimentally observed \cite{e2,e3}. Thus, assuming only rational $\hbar _{\mathrm{s}}=l/q$ ($l$ and $q$ are coprime integers), the QE levels (\ref{wj}) for $j=gl$ ($g$ arbitrary integer) are given by 
\begin{equation}
\omega _{gl}=\omega +2\pi gl(\beta +q/2)\ \ \mathrm{mod}(2\pi )  \label{wg}
\end{equation}
and, due to (\ref{bj}), they are all associated with a fixed value of $\beta 
$, $\beta _{gl}=\beta $. For irrational $\beta$, the ladder of equally-spaced levels (\ref{wg}) covers densely the entire QE range. For rational $\beta $ (strict quantum resonance), this ladder consists of a finite number $\bar{g}$ of levels with spacing $\Delta \omega =2\pi /\bar{g}$, where $\bar{g}$ is the smallest integer such that $\bar{g} l(\beta +q/2)$ is integer; see, e.g., Fig. 2(b) for $\beta =1/5$.

For $\beta =0$ (usual kicked rotor) there are either no spectral ladders ($\bar{g}=1$ for $q$ even) or trivial ladders with spacing $\Delta\omega =\pi$ ($\bar{g}=2$ for $q$ odd), so that the spectra exhibit almost no ladder regularity, see Fig. 2(a).

The case of the main quantum resonances, $\hbar _{\mathrm{s}}=l$ ($q=1$), is exactly related \cite{qr} to an integrable version of (\ref{KP}), the linear kicked rotor \cite{ali}. This relation implies that for $\hbar _{\mathrm{s}}=l$ and generic irrational $\beta$ dynamical localization takes place and the entire QE spectrum of the $\beta$-kicked-rotor is given by Eq. (\ref{wg}) with $\omega =k\int_0^{2\pi}V(x)dx/(2\pi )$. 

We now consider quantum-dynamical manifestations of the QE regularities. An arbitrary initial wavepacket associated with a regular subspectrum (\ref{wj}) is a general linear combination of the particle QE states (\ref{tqe}): 
\begin{equation}
\phi _{\beta ,\omega }(x)=\exp (i\beta x)\psi _{\beta ,\omega }(x)\bar{\chi} (x),
\label{aiwp}
\end{equation}
where $\bar{\chi} (x)$ is a periodic function with period $\hbar $,
\begin{equation}
\bar{\chi} (x)=\sum_{j=-\infty }^{\infty }c_{j}\exp (2\pi ijx/\hbar ),  \label{chi}
\end{equation}
$c_{j}$ being arbitrary coefficients. Using (\ref{qee}) and (\ref{wj}), we obtain the wavepacket after $s$ kicks:
\begin{eqnarray}
\hat{U}^{s}\phi _{\beta ,\omega }(x) &=& e^{i\beta x}\psi _{\beta ,\omega }(x)\sum_{j=-\infty }^{\infty }c_{j}e^{2\pi ijx/\hbar -i\omega_j s} \notag \\
&=& \exp [i(\beta x-\omega s)]\psi _{\beta ,\omega }(x)\chi _{s}(x),  \label{awp}
\end{eqnarray}
\begin{equation}
\chi _{s}(x)=\sum_{j=-\infty }^{\infty }c_{j}\exp \{ 2\pi ij[x/\hbar -(\beta
+\pi j/\hbar )s]\}.  \label{chis}
\end{equation}
The essential time ($s$) dependence of (\ref{awp}) is due to the function (\ref{chis}). The latter evolves \emph{freely} with frequencies $2\pi j\beta+\pi j^2/\hbar _{\mathrm{s}}$, independent of the nonintegrability, and is quasiperiodic in $s$ for irrational values of $\hbar _{\mathrm{s}}$ and/or $\beta$. 

For rational $\hbar _{\mathrm{s}}=l/q$, the result (\ref{chis}) assumes an interesting simple form, reflecting the ladder regularity, if $j$ is restricted to multiples $g$ of $l$, $j=gl$, as above: 
\begin{eqnarray}
\chi _{s}(x) &=&\sum_{g=-\infty }^{\infty }c_{gl}\exp [2\pi igl(x/\hbar
-\beta ^{\prime }s)]  \notag \\
&=&\chi _{0}(x-\beta ^{\prime }\hbar s),  \label{chisqr}
\end{eqnarray}
where $\beta ^{\prime }=\beta +q/2\ \mathrm{mod}(1)$. Eq. (\ref{chisqr})
means that the component $\chi _s(x)$ of (\ref{awp}) is a \emph{traveling wave} moving without change of shape at constant velocity $\beta ^{\prime}\hbar$. The time evolution of $|\chi_s(x)|^2$ for both irrational and rational values of $\hbar _{\mathrm{s}}$ is illustrated in Fig. 3.

It is clear from Eq. (\ref{awp}) that the expectation value of any physical observable in a general state (\ref{awp}) evolves only with the free frequencies $\omega_j-\omega_{j^{\prime}}$, associated with matrix elements of the observable between translated eigenstates (\ref{tqe}). In this sense, one has a free quantum dynamics for arbitrarily strong chaos.
\begin{figure}[tbp]
\includegraphics[width=8.7cm]{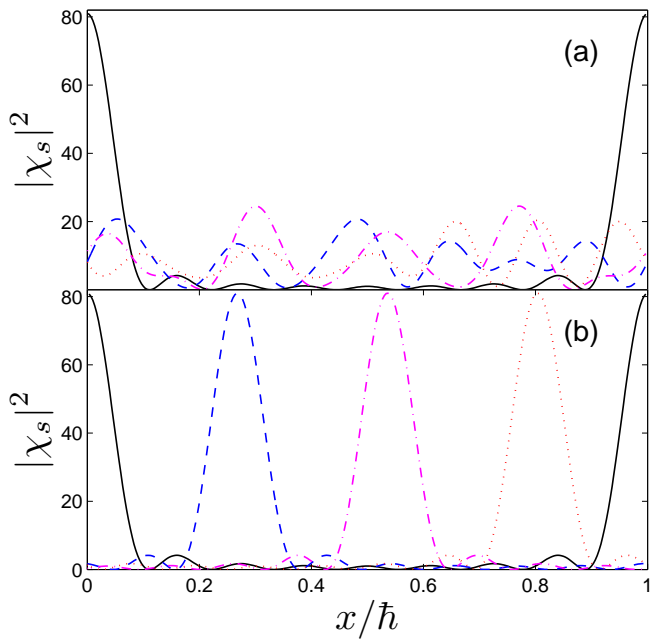}
\caption{(Color online) (a) Time ($s$) evolution of $|\chi_{s}(x)|^{2}$, where $\chi _{s}(x)$ is given by Eq. (\ref{chis}) with $c_{j}=1$ for $|j|\leq 4$ and $c_{j}=0$ otherwise, for $\beta =2-\sqrt{3}$, $\hbar _{\mathrm{s}}=(\sqrt{5}
-1)/2 $, and: $s=0$ (solid line), $s=1$ (blue dashed line), $s=2$
(magenta dot-dashed line), and $s=3$ (red dotted line). (b) Similar to (a)
but for the quantum-resonance case of $\hbar _{\mathrm{s}}=1/2$; clearly, 
$|\chi _{s}(x)|^{2}$ evolves in this case as a traveling wave with
constant shape and velocity, see Eq. (\ref{chisqr}).}
\label{fig3}
\end{figure}

At least in principle, a good approximation to an initial wavepacket (\ref{aiwp}) may be prepared by exposing the quantum particle, initially in some momentum state $p=\bar{p}$ with $\bar{p}/\hbar<1$, to a series of kicks with potential having spatial period $\hbar$ (instead of $2\pi$). After a large number of sufficiently strong kicks, the particle will be in the state $\varphi_{\beta}(x)=\exp (i\beta x)\bar{\chi} (x)$, where $\beta =\bar{p}/\hbar$ and $\bar{\chi} (x)$ is given by (\ref{chi}) with many significant harmonics $c_j$. The state $\varphi_{\beta}(x)$ corresponds to (\ref{aiwp}) with $\psi _{\beta ,\omega }(x)=1$, a zero-momentum state. The latter is a good approximation to a QE state of the $\beta$-kicked-rotor for generic irrational $\beta$ and $k\ll 1$ in (\ref{KP}), due to dynamical localization with a very small localization length. Then, the time evolution of $|\varphi_{\beta ,s}(x)|^2$ is well approximated by that of $|\chi _s(x)|^2$.    

In conclusion, we have considered the QE spectra of the paradigmatic nonintegrable system (\ref{KP}) for all values of an experimentally relevant quantity, the quasimomentum $\beta$. We have shown that these spectra generically exhibit new kinds of significant regularities which persist independently of the chaos strength. This is in contrast with usual spectral regularities \cite{br} which disappear in fully chaotic regimes. The generic regularities are dense subspectra of either ``Poisson" type (\ref{wj}) for irrational $\hbar _{\mathrm{s}}$ or ladder type (\ref{wg}) for rational $\hbar _{\mathrm{s}}$. These subspectra emerge from features of the free motion between kicks that are unaffected by an arbitrary periodic potential $V(\hat{x})$ in (\ref{KP}). Since this basic origin of the regular subspectra is present also in more general kicked-particle systems \cite{e1,e6}, our results can be extended to these systems.

The usual kicked rotor, studied in many works, corresponds to the case $\beta =0$ of the kicked particle. Our results show that this case is special for rational $\hbar _{\mathrm{s}}$, since the ladder regularity for generic $\beta$ is essentially absent for $\beta =0$, see Fig. 2. For generic (irrational) $\hbar _{\mathrm{s}}$, however, there is no such difference between $\beta =0$ and other values of $\beta$, since the ladder regularity is absent for all $\beta$, being replaced by the Poisson one.

A most interesting manifestation of the QE spectral regularities is that the quantum dynamics associated with a regular subspectrum exhibits only free-motion frequencies for arbitrarily strong chaos. This free quantum dynamics and related phenomena in wavepacket evolution, such as the traveling waves (\ref{chisqr}) (see Fig. 3(b)), should be experimentally observable at least in simple cases, using, e.g., the procedure above for preparing an approximate initial state (\ref{aiwp}).
 
This work was partially supported by Bar-Ilan University Grant No. 2046.


\begin{thebibliography}{99}
\bibitem{note0} If both $\hbar _{\mathrm{s}}$ and $\beta$ are rational (strict quantum resonance), the QE spectrum consists of a finite number of levels at any fixed value of a conserved ``quasiangle" $\alpha$. As $\alpha$ is varied, each level broadens into a band. We assume here fixed $\alpha$, but in some cases the bands are ``flat", i.e., infinitely degenerate levels. See more details in Ref. \cite{qr}.
\bibitem{note} After inserting (\ref{bf}) into Eq. (\ref{qee}), we easily find that $\psi _{\beta ,\omega }(x)$ is an eigenstate of the operator $\hat{U}_{\beta}=\exp [-i(\hat{p}+\beta\hbar )^{2}/(2\hbar )]\exp [-ikV(\hat{x})/\hbar ]$ with eigenvalue $\exp(-i\omega)$, where one has to interpret now $\hat{p}$ as an angular-momentum operator with eigenvalues $n\hbar$ ($n$ integer) since $\psi_{\beta ,\omega }(x)$ is $2\pi$-periodic. The operator $\hat{U}_{\beta}$ defines then a $\beta $-kicked-rotor with angle $x$; the usual kicked rotor corresponds to $\beta =0$. One can also interpret $\hat{p}+\beta\hbar$ in $\hat{U}_{\beta}$ as the original (linear) momentum, with $\beta$ being its ``fractional" part in units of $\hbar$ \cite{kp}.
\end{thebibliography}
\end{document}